\documentclass[aip,pop]{revtex4-1}

\usepackage{graphicx}
\usepackage{subfigure}
\usepackage{epstopdf}
\usepackage{amsmath}
\usepackage{dcolumn}
\usepackage{bm}
\usepackage[colorlinks=true, linkcolor=blue]{hyperref}

\begin{document}

\title{Practicality of magnetic compression for plasma density control}
\author{Renaud Gueroult}
\affiliation{Princeton Plasma Physics Laboratory, Princeton University, Princeton, NJ 08543 USA}
\author{Nathaniel J. Fisch}
\affiliation{Princeton Plasma Physics Laboratory, Princeton University, Princeton, NJ 08543 USA}

\begin{abstract}
Plasma densification through magnetic compression has been suggested for time-resolved control of the wave properties in plasma-based accelerators~\cite{Schmit2012}. Using particle in cell simulations with real mass ratio, the practicality of large magnetic compression on timescales shorter than the ion gyro-period is investigated. For compression times shorter than the transit time of a compressional Alfven wave across the plasma slab, results show the formation of two counter-propagating shock waves, leading to a highly non-uniform plasma density profile. Furthermore, the plasma slab displays large hydromagnetic like oscillations after the driving field has reached steady state. Peak compression is obtained when the two shocks collide in the mid-plane. At this instant, very large plasma heating is observed, and the plasma $\beta$ is estimated to be about $1$. Although these results point out a densification mechanism quite different and more complex than initially envisioned, these features still might be advantageous in particle accelerators. 
\end{abstract}

\maketitle
\endpage{}

\section{Introduction}

One of the main phenomena limiting the electron energy in plasma-based particle accelerators is electron dephasing (see, for \emph{e.g.}, Ref.~\cite{Esarey2009}). Assuming a constant phase velocity $v_p<c$, with $c$ the speed of light, for the plasma wave, the velocity $v_z$ of an electron accelerated along the $z$ axis eventually becomes greater than the wave phase velocity. When $v_z > v_p$, the electron outruns the wave, and it moves into a decelerating phase region of the plasma wave. 

Using magnetic compression to control the time-resolved plasma wave properties was recently suggested~\cite{Schmit2012} as a possible option to overcome electron dephasing. In the proposed method, the plasma density is modulated using a time-dependent externally applied magnetic field, so that the phase velocity $v_p$ can match the electron velocity over a larger spatial region. The physical mechanism proposed for the plasma densification is the following: within a bounded parameter regime, the induced azimuthal electric field $E_{\phi} = -r\dot{B}/2$ associated with the time variation of the uniform external axial magnetic field $B(t)$ causes an inward radial drift of both electrons and ions. The resulting plasma densification is such that $n \sim B(t)$. In addition, this densification mechanism is predicted to hold even for timescales short compared to the ion gyro-period $\tau_{ci} = 2\pi/\omega_{ci}$, with $\omega_{ci}$ the ion gyro-frequency, thanks to an averaging over the gyrophases continuum. 

In essence, the compression scheme proposed in Ref.~\cite{Schmit2012} is a fast parallel theta-pinch setup. In theta-pinch experiments, compression and plasma heating are classically described by a "bounce"~\cite{Morse1967,Dove1971} or a "snowplow"~\cite{Rosenbluth1954, Uchida1962} model depending on the experiment parameters. The bounce model has been applied for collisionless plasmas, such as predicted for the application considered here. For a sufficiently large compression ratio, a collisionless shock is associated with the compression when $\tau_r \ll \tau_{ii}$ and $\tau_r \leq \Delta x/V_A$, where $\tau_r$ and $\tau_{ii}$ are respectively the current rise-time in the theta-coil and the ion-ion collision time, $\Delta x$ is the plasma width along the compression direction (radius in a theta pinch), and $V_A$ is the Alfven velocity.     

Since the parameters regime envisioned for plasma densification through magnetic compression differs from the typical theta-pinch operating conditions, direct transposition of these results is difficult. However, in light of these results, it seems that the densification mechanism proposed in Ref.~\cite{Schmit2012} is overly simplified. In order to confirm the potential of magnetic compression for plasma densification, a better understanding of the physics controlling the compression process in this regime is required. 

In this paper, we investigate, through particle-in-cell simulations, the practicality of plasma magnetic compression for time-dependent wave properties control. The parameter regime studied here is suitable for electron dephasing in plasma-based particle accelerators. In Sec.~\ref{Sec:II}, the configuration is introduced. In Sec.~\ref{Sec:III}, global results illustrating the evolution of plasma parameters particularly relevant to wave properties control are presented, and apparent limitations are highlighted. In Sec.~\ref{Sec:IV}, the initial plasma compression phase is analyzed, and the shock dynamics leading to density non-uniformity is exposed. In Sec.~\ref{Sec:V}, plasma properties near peak compression and plasma expansion are discussed. In Sec.~\ref{Sec:Summary}, the main findings are summarized.

\section{Problem description}

\label{Sec:II}

Although typical accelerator channels are cylindrical, the configuration chosen here, for simplicity, is a 1D plasma slab. Due to this simplification, geometrical densification effects are lost. As a result, the compression ratio at peak compression differs from what would be the ratio of a cylindrical plasma column for the same compression parameters.  On the other hand, this choice greatly simplifies the analysis of the propagation and reflection of the compression waves, while conserving all the compression physics.

%The domain includes a vacuum region on the left hand side, and a electron-proton plasma region on the right hand side. The plasma interface is located at $x = 0.8L$, where $L$ is the length of the computational domain. The physical width of the plasma region is $1~$cm.

\begin{figure}
\begin{center}
\includegraphics[]{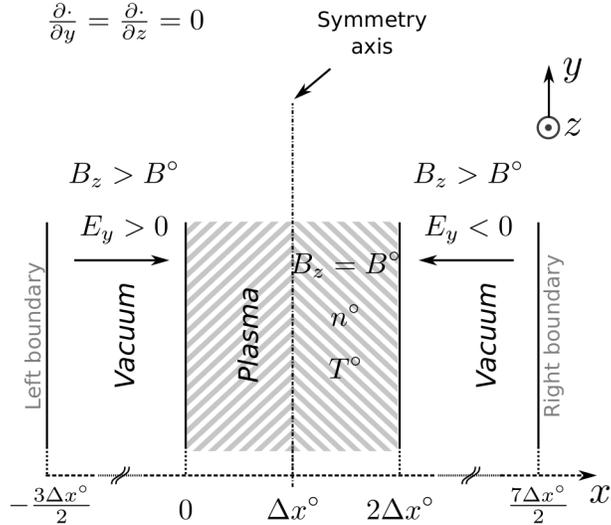}
\label{Fig:Sketch}
\caption{Computational domain. The domain includes a plasma region surrounded by a vacuum region on each side of the plasma. Both the plasma and vacuum regions are initially immersed in a uniform bias magnetic $\mathbf{B} = {B}^{\circ}\hat{z}$. Magnetic compression ramps are generated at each boundary of the domain and propagate toward the plasma region. Boundaries transmit outgoing waves reflected by the plasma. }
\end{center}
\end{figure}

The configuration studied is depicted in Fig.~\ref{Fig:Sketch}.  The initial value of parameter $p$ is denoted $p^{\circ}$. The domain includes an electron-proton plasma region in the center, surrounded by a vacuum region on each side. The width of the plasma slab is $2{\Delta x}^{\circ} = 20~$mm, which is about $375$ electron skin depths for the plasma parameters chosen (see Table~\ref{Table:initial_conditions}). The vacuum region extends for $3{\Delta x}^{\circ}/2$ on each side of the plasma. Both the plasma and vacuum regions are initially immersed in a uniform bias magnetic field $\mathbf{B}(x,t=0) = {B}^{\circ}~\hat{z}$, with ${B}^{\circ} = 5~$T. The initial plasma density is ${n_e}^{\circ} = {n_i}^{\circ} = {n}^{\circ} = 10^{16}~$cm$^{-3}$, so that the plasma is magnetically overdense with ${\omega_{pe}}^{\circ}/{\omega_{ce}}^{\circ}\gtrsim 6$. The ion and electron temperatures are ${T_e}^{\circ} = {T_i}^{\circ} = {T}^{\circ} = 20$~eV. The initial kinetic to magnetic pressure ratio ${\beta}^{\circ} = 2\mu_0 {n}^{\circ} k_B {T}^{\circ}/{{B}^{\circ}}^2 \sim 2~10^{-3}$, and grows to $\beta \sim \mathcal{O} (1)$ as the temperature increases during the plasma compression.

\begin{table}
\begin{center}
\begin{tabular}{c | c}
Parameter & Value\\
\hline
Plasma slab width $2{\Delta x}^{\circ}$ [mm] & $20$\\
Electron and ion density ${n_e}^{\circ}$ and ${n_i}^{\circ}$ [cm$^{-3}$] & $10^{16}$\\ 
Electron and ion temperature ${T_e}^{\circ}$ and ${T_i}^{\circ}$ [eV] & $20$\\ 
Bias (background) magnetic field ${B}^{\circ}$ [T] & $5$\\
Plasma frequency ${\omega_{pe}}^{\circ}$ [s$^{-1}$] & $5.6~10^{12}$\\
Electron gyro-frequency ${\omega_{ce}}^{\circ}$ [s$^{-1}$] & $8.8~10^{11}$\\
Ion gyro-frequency ${\omega_{ci}}^{\circ}$ [s$^{-1}$] & $4.8~10^{8}$\\
Debye length ${\lambda_D}^{\circ}$ [$\mu$m]&  $0.3$ \\
Electron skin depth ${c/\omega_{pe}}^{\circ}$ [$\mu$m]& $50$\\
Alfven velocity ${v_A}^{\circ}$ [m.s$^{-1}$] & $1.1~10^{6}$\\
Sound speed ${c_s}^{\circ}$ [m.s$^{-1}$] & $\sim 10^{5}$\\
Electron gyro-radius ${\rho_{L_e}}^{\circ}$ [$\mu$m] & $3$\\
Ion gyro-radius ${\rho_{L_i}}^{\circ}$ [$\mu$m]& $130$\\
Plasma ${\beta}^{\circ}$ & $2~10^{-3}$\\
\end{tabular}
\caption{Initial plasma parameters at $t=0$. The notation $p^{\circ}$ is used to denote $p|_{t=0}$. }
\label{Table:initial_conditions}
\end{center}
\end{table}

Plasma dynamics is modeled using the one-dimensional version of the fully electromagnetic and relativistic particle in cell code \textsc{Epoch}~\cite{Arber2015}. Particles are followed in one spatial dimension ($x$) and three velocity dimensions. Due to the slab assumption, only three field components are retained, with $\mathbf{E} = (E_x,E_y,0)$ and $\mathbf{B} = (0,0,B_z)$. The size of a grid cell is defined as the Debye length $\lambda_D$, and $30-50$ particles per cell are used. The time step is chosen to satisfy the Courant condition. Real proton and electron masses are used ($m_p/m_e\sim1846$).

On the left hand side boundary, right propagating electromagnetic waves are launched, whereas left propagating waves are transmitted without reflection. Boundary condition for particles on this boundary has no influence since no particle ever reaches it.  The transverse electric field $E_y$ is specified on the left boundary so that the magnetic field component of the right propagating wave is 
\begin{equation}
B_z\left(x = -\frac{3{\Delta x}^{\circ}}{2},t\right) = 
\begin{cases}
    {B}^{\circ} \left[1+\delta_B\sin\left(\frac{ \pi t}{ 2\tau_r}\right)\right] & \textrm{for} \quad t\leq\tau_r\\
    {B}^{\circ} \left[1+\delta_B\sin\left(\frac{ \pi t}{ 2\tau_s}+\frac{ \pi}{ 2}\right)\right] & \textrm{for} \quad t\geq\tau_r\\
\end{cases}
\label{Eq:BField}
\end{equation}
where $\tau_r$ is the driving magnetic field rise-time, $\tau_m$ is the time over which the driving field is maintained and $\delta_B$ is the driving to bias magnetic field ratio. For $t\geq\tau_r$, the amplitude of the magnetic field component of the right propagating wave at the left boundary is nearly constant. Following Ref.~\cite{Schmit2012}, the compression parameters are $\tau_r = 5$~ns and $\delta_B = 2$, and $\tau_m$ is taken as $10^3\tau_r$ which ensures an almost constant driving field for $t>\tau_r$.

On the right hand side boundary, left propagating waves are launched, while right propagating waves are transmitted without reflection. Similarly, boundary condition for particles at this boundary has no influence. The transverse electric field at the right boundary is in phase opposition with the one at the left boundary, so that the magnetic field component of the left propagating wave is identical to the one given in Eq.~\ref{Eq:BField}, \emph{i. e.}  $B_z(-3{\Delta x}^{\circ}/2,t) = B_z(7{\Delta x}^{\circ}/2,t)$. 

%\textcolor{cyan}{Need to add to this paragraph if we decide to leave it as part of this manuscript} Because of the symmetry of the domain, one could in principle only model half of the domain, and apply a symmetry boundary condition for both particle and fields in the mid-plane. However, the use of a perfectly reflecting and conducting boundary in the mid-plane, sometimes referred to as the \emph{piston method} (see, \emph{e. g.}, Ref.~\cite{Burgess1989}) and routinely adopted to simulate counter-streaming plasmas and collisionless shock formation, yields results different than the those from full domain simulations.

%On the right hand side boundary, a perfectly reflecting and conducting boundary is imposed. This method, sometimes referred to as the \emph{piston method} (see Ref.~\cite{Burgess1989}), is routinely used to simulate counter-streaming plasmas and collisonless shock formation, and allows to reduce the simulation domain by half. To ensure with full domain simulations, simulations on the full domain were conducted and compared to the results obtained for the half domain with the reflecting wall.

\begin{table}
\begin{center}
\begin{tabular}{c | c}
Parameter & Value\\
\hline
Driving to bias magnetic field ratio $\delta_B$ & $2$\\
Magnetic field rise-time $\tau_r$ [ns] & $5$\\
$\tau_r {\omega_{ci}}^{\circ}$ & $2.4$\\
\end{tabular}
\caption{Compression parameters.}
\label{Table:compression_parameters}
\end{center}
\end{table}

The right (\emph{resp.} left) propagating compression wave generated at the left (\emph{resp.} right) boundary propagates towards the plasma boundary at the speed of light. Although the frequency associated with the driving field is much lower than the plasma frequency $\omega_{pe}$, part of the driving field is transmitted to the plasma due to the plasma pre-magnetization. Introducing $\mu-i\chi = [({\epsilon_{\perp}}^2-{\epsilon_{\times}}^2)/\epsilon_{\perp}]^{1/2}$, with  $\epsilon_{\perp}$ and $\epsilon_{\times}$ respectively the perpendicular and cross-field component of the dielectric tensor, the power transmission coefficient through the vacuum-plasma interface reads 
\begin{equation}
\mathcal{T} = \frac{4\mu}{(1+\mu)^2+\chi^2}.
\end{equation}
Using the collisionless cold plasma dispersion relation for a fast magneto-sonic wave, one finds that for $\omega_{pe}/\omega_{ce}\sim6$, $\mathcal{T} \sim 1.5\%$. Because of this very small transmission coefficient, this model predicts a reflected wave of amplitude nearly equal to the incident wave. For this reason, and although the large amplitude compression setup considered here differs greatly from this idealized linear wave picture, large power reflection is anticipated at the vacuum-plasma interface.

\section{Global results}

\label{Sec:III}

\begin{figure}
\begin{center}
\includegraphics[]{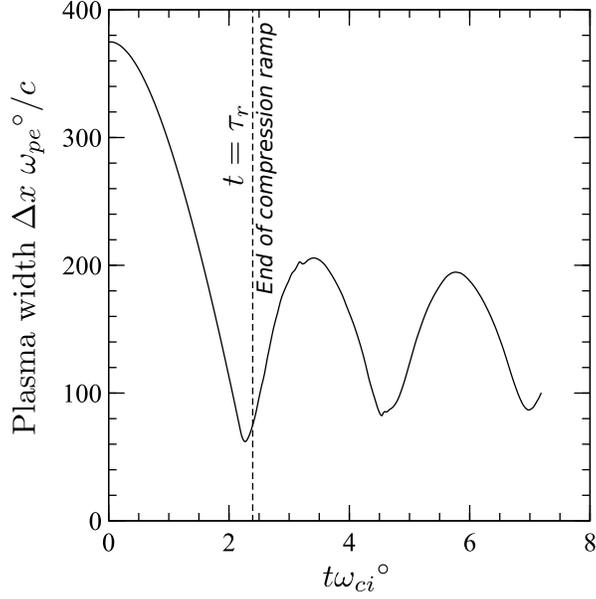}
\caption{Time evolution of the plasma slab width $\Delta x$. Peak compression is obtained right before the driving field reaches its maximum value ($t = \tau_r$). Past this instant, the plasma slab shows successive expansion and compression phases, with weak damping. }
\label{Fig:Compression}
\end{center}
\end{figure}

The evolution of the plasma slab width $\Delta x$ with time is plotted in Fig.~\ref{Fig:Compression}. Peak compression, which is defined as the instant for which $\Delta x/{\Delta x}^{\circ}$ is minimum, occurs in this setup for $t{\omega_{ci}}^{\circ}=2.26$, that is to say right before the driving field $B$ reaches its maximum  ($\tau_r{\omega_{ci}}^{\circ} = 2.39$).  Following peak compression, the plasma experiences successive compression and expansion phases. These oscillations appear similar to hydromagnetic oscillations~\cite{Niblett1959}, with a period $\tau \sim {\Delta x}^{\circ}/v_A\sim1/{\omega_{ci}}^{\circ}$, where $v_A$ is the Alfven velocity calculated for the maximum driving field $B$ and ${\omega_{ci}}^{\circ}$ is the ion cyclotron angular frequency. This result is consistent with oscillations observed in fast-rising theta-pinches~\cite{Fisher1962,Martone1971}. 

\begin{figure}
\begin{center}
\subfigure[)~Ion density $n_i$ ]{\includegraphics[]{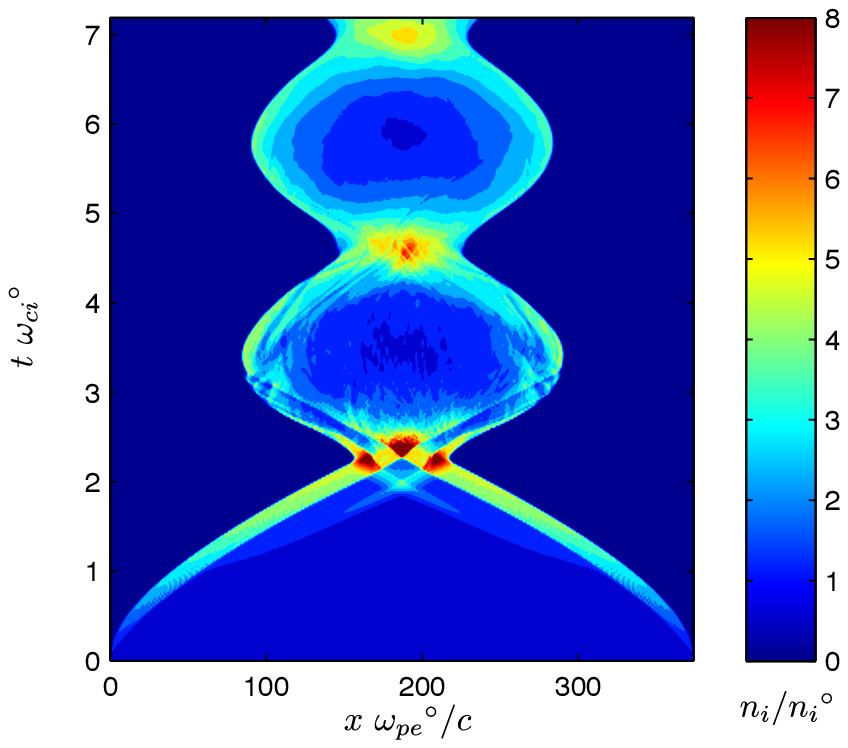}\label{Fig:Ni}}\subfigure[)~Magnetic field $B$ ]{\includegraphics[]{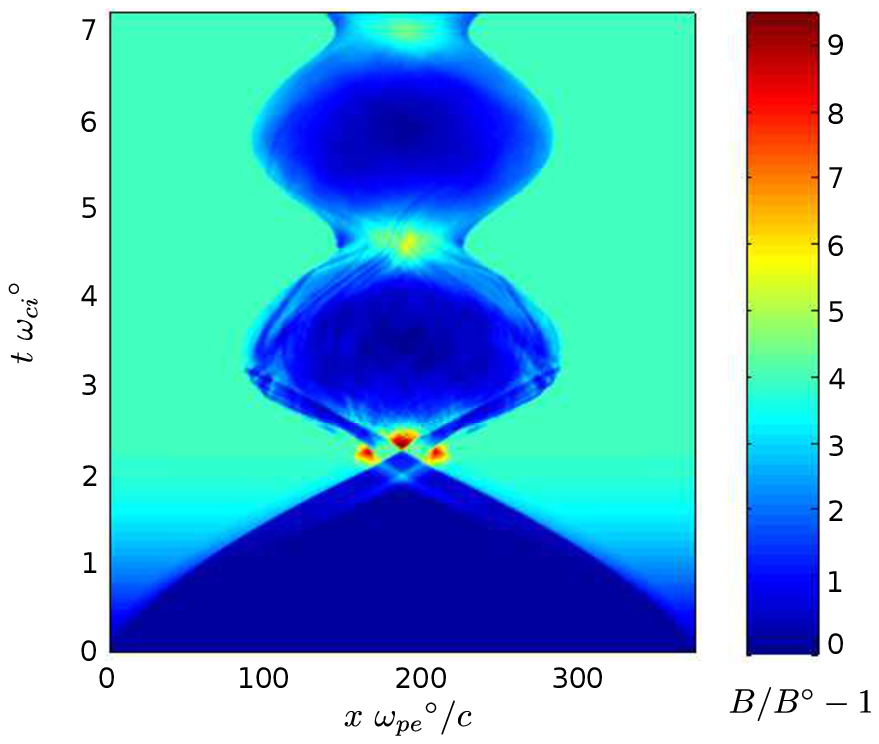}\label{Fig:B}}\\\subfigure[)~Ion temperature $T_i$]{\includegraphics[]{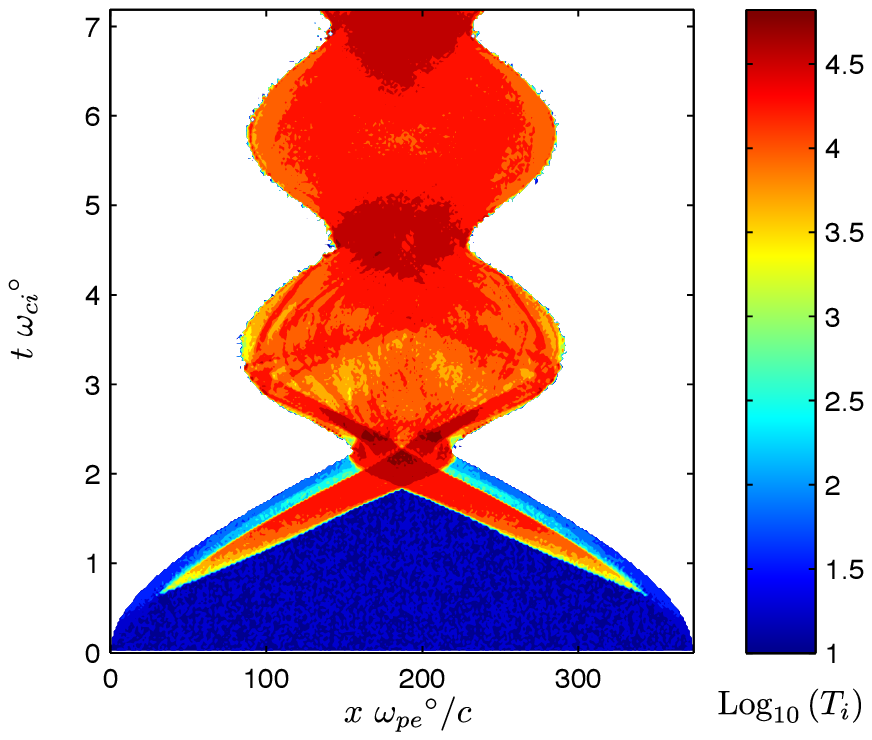}\label{Fig:Ti}}\subfigure[)~Electron temperature $T_e$]{\includegraphics[]{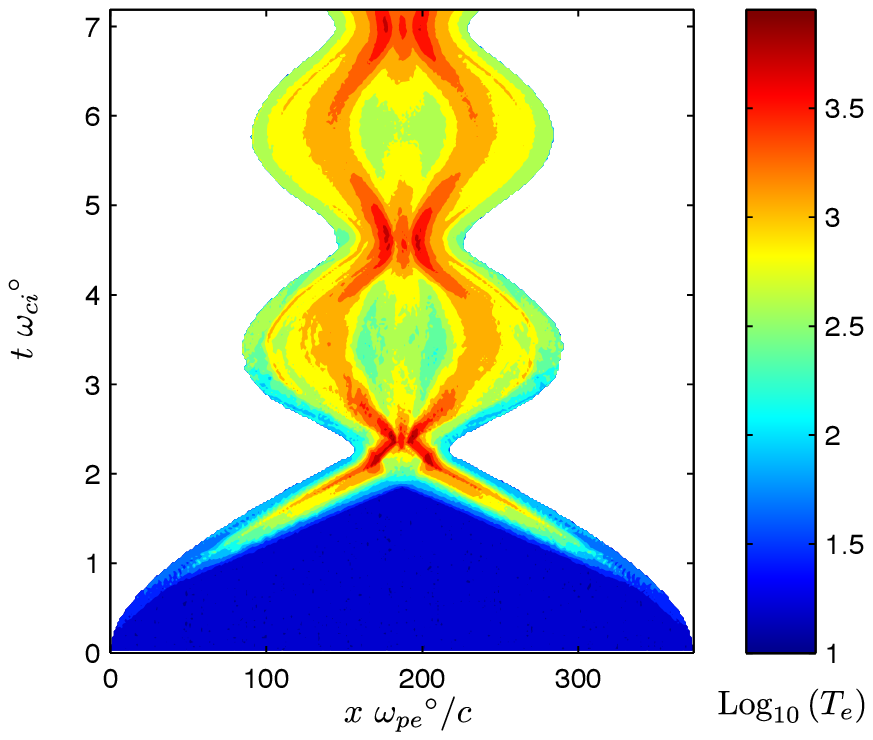}\label{Fig:Te}}
\caption{Evolution of the plasma parameters in response to the compressing wave: ion number density~(\subref{Fig:Ni}), magnetic field~(\subref{Fig:B}), ion temperature~(\subref{Fig:Ti}) and electron temperature~(\subref{Fig:Te}). Initial plasma parameters are listed in \protect{Table~\ref{Table:initial_conditions}}.}
\label{Fig:Full}
\end{center}
\end{figure}

Figure~\ref{Fig:Full} presents the simulated time evolution of the plasma parameters in the spatial domain corresponding to the initial plasma slab width ${\Delta x}^{\circ}\sim 375~c/{\omega_{pe}}^{\circ}$. Looking at the plasma density profile in Fig.~\ref{Fig:Ni}, one immediately notices that the density is highly non-uniform across the plasma slab. More precisely, the simulations depict a plasma density profile that is almost always hollow, with the exception of peak compression times ($t{\omega_{ci}}^{\circ}\sim2.26$, $4.54$ and $6.99$), for which the density appears to peak in the mid-plane. As the compression and expansion phases follow each other, the density gradient scale-length appears to grow, making density gradients across the plasma slab less severe.

The trends observed on the plasma density are confirmed when looking at the magnetic field profile plotted in Fig.~\ref{Fig:B}. With the exception of peak compression times, the magnetic field profile is hollow, and shows strong variations in a region near the plasma-vacuum interface. As anticipated, the field in the vacuum region is significantly larger than the driving field amplitude due to reflection at the plasma-vacuum interface. For a fully reflected driving field, one expects a field amplitude $(1+2\delta_B){B}^{\circ}$ in the vacuum region, which seems consistent with the $B/{B}^{\circ}\lesssim5$ observed in Fig.~\ref{Fig:B}. Furthermore, qualitative comparison of Fig.~\ref{Fig:Ni} and \ref{Fig:B} suggests limited variations of $B/n$ during compression.

The ion and electron temperature evolution is depicted in Fig.~\ref{Fig:Ti} and Fig.~\ref{Fig:Te}, showing very large ion heating as well as significant electron heating. Ion and electron temperature increases $T_i/{T_i}^{\circ}$ and $T_e/{T_e}^{\circ}$ in excess of respectively $2000$ and $150$ are locally measured. Assuming for the sake of simplicity that $B/n$ remains constant at peak compression, the ion temperature required to balance the magnetic pressure $\mathcal{P_B} = B^2/(2\mu_0)$ is $T_i \sim {T_i}^{\circ} B/({\beta}^{\circ}{B}^{\circ})$. For ${\beta}^{\circ}\sim2~10^{-3}$ (see Table~\ref{Table:initial_conditions}), and $B/{B}^{\circ} = 5$, the required plasma heating is $T_i/{T_i}^{\circ}\sim 2300$. Since these ion temperatures are very close to the ones observed near peak compression, one can infer that $\beta = nkT/\mathcal{P_B}\sim\mathcal{O}(1)$ for peak compression.

Similarly to what was observed for the ion density in Fig.~\ref{Fig:Ni}, temperature gradients across the plasma slab weaken with time, with $T_i$ becoming rather uniform for $t{\omega_{ci}}^{\circ}\geq5$. Although a similar uniformization trend is observed for $T_e$,  relatively strong non-uniformity prevails at longer times. However, it stands to reason that collisional effects, which are neglected in this study, would lead at later times to electron heating and thermalization. For the initial plasma parameters considered here (see Table~\ref{Table:initial_conditions}), the ion and electron collision frequency are respectively ${\nu_i}^{\circ}\sim{\omega_{ci}}^{\circ}/5$ and ${\nu_e}^{\circ}\sim{\omega_{ci}}^{\circ}/135$. Although heating effects associated with compression could extend significantly the duration over which collisional effects are negligible, these estimates suggest $t{\omega_{ci}}^{\circ}\sim5$ as a conservative upper limit for collisionless modeling.

These global numerical results suggest that plasma densification through magnetic compression, such as proposed in Ref.~\cite{Schmit2012}, will not occur as envisioned. Numerical results highlight two main main hurdles towards this goal.  First, plasma parameters, in particular density,  show significant variations across the plasma slab. These density gradients are associated with magnetic field gradient across the plasma, which differs from the homogeneous magnetic field amplification described in Ref.~\cite{Schmit2012}.  Second, the plasma slab width displays strong oscillations, rather than converging towards a steady-state solution as the driving field reaches its set value. This occurs even on timescales short compared to the collision time. On the other hand, one might be able to take advantage of these features, in particular the profile hollowness, to better focus a particle beam in a particle accelerator. In order better to appreciate whether these phenomena are true show-stoppers for the use of magnetic compression for wave properties control in plasma based particle accelerators, or simply added complexity,  the physics of the initial compression phase and of the peak compression and following expansion is analyzed.

\section{Initial compression phase}

\label{Sec:IV}

In this section, only the results obtained for the left half of the domain are presented since the other half is a mirror image.

\begin{figure}
\begin{center}
\includegraphics[]{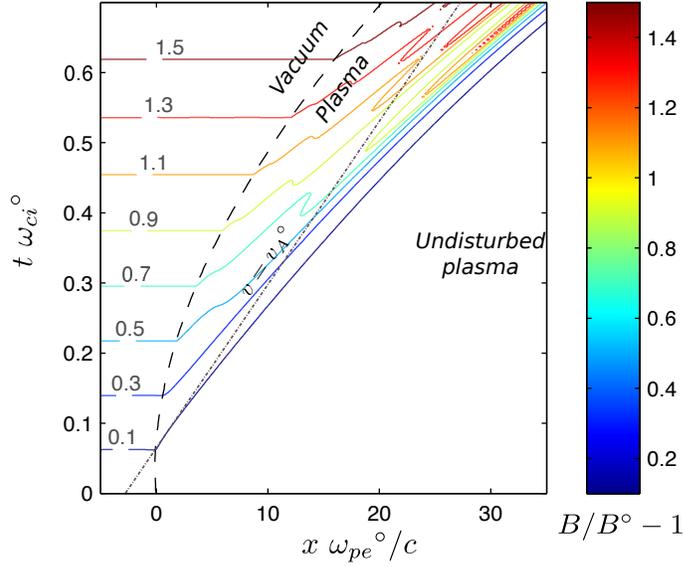}
\caption{Contour of $B/{B}^{\circ}-1$ at early times. The black dashed line represents the plasma-vacuum boundary. The grey dashed-dotted line indicates propagation at the initial Alfven velocity ${v_A}^{\circ} = {B}^{\circ}/\sqrt{\mu_0 {n}^{\circ}m_p}$. For low driving fields ($B/{B}^{\circ}-1\ll1$), the propagation velocity is close to ${v_A}^{\circ}$. As the driving field amplitude increases ($B/{B}^{\circ}-1\sim\mathcal{O}(1)$), a trailing wave train is observed in the piston region.}
\label{Fig:BFieldContours}
\end{center}
\end{figure}

%Since the wave transit time across the plasma slab $\widetilde{\Delta x}/\widetilde{v_A}\sim 10$~ns is comparable to the magnetic field %rise time $\tau_r = 5~$ns, one expects a non uniform axial profile for both the density and the magnetic field as the driving field %progressively compresses the plasma. 
At early times,  the driving field amplitude at the vacuum-plasma boundary is much smaller than the bias field ( $|B/{B}^{\circ}-1|\ll 1$). 
As shown in Fig.~\ref{Fig:BFieldContours}, the magnetic field perturbation propagates in the plasma with a velocity $v\sim v_A = B/\sqrt{\mu_0 nm_p}$, with $m_p$ the proton mass. This is consistent with the phase velocity $v_{ms}$ of the fast magneto-sonic wave since the sound speed ${c_s}^{\circ}\sim 0.1 {v_A}^{\circ}$, and hence $v_{ms} = ({{v_A}^{\circ}}^2+{{c_s}^{\circ}}^2)^{1/2}\sim{v_A}^{\circ}$. In first approximation, $B/n$ can be assumed constant, and $v_A \propto \sqrt{n} \propto \sqrt{B}$. Consequently, the phase velocity of the wave increases as the perturbation becomes larger, and the wave front steepens. This is illustrated by the decrease of the slope between the first and second contour in Fig.~\ref{Fig:BFieldContours}.

Since the time of passage of the magnetic disturbance is small compared to the ion gyro-period ($\omega_{ci}\ll 2\pi/\tau_r\ll\omega_{ce}$), electrons and ions response to this disturbance differs: electrons are magnetized, whereas ions are essentially non-magnetized. For small enough perturbations, electrons motion, which consist mostly in the $\mathbf{E} \times \mathbf{B}$ drift, is essentially in the $x$ direction, with velocity $E_y/B$. In the absence of a longitudinal electric field, ions motion would be limited to a velocity kick in the $y$ direction. Because of these different dynamics, a charge separation occurs, and a longitudinal electric field is formed to ensure that ions and electrons exhibit the same displacement in the $x$ direction after the pulse passage. An estimate for the amplitude of this longitudinal field can be obtained from the cold plasma solution of the fast magneto-sonic mode. The longitudinal to total electric field ratio goes from $0$ for $\omega\ll\omega_{ci}$ to $1$ as $\omega$ approaches the resonance frequency $\omega_{lh}$, with $\omega_{lh}$ the lower hybrid frequency. Quantitatively, $E_y/E_x = \epsilon_{\times}/\epsilon_{\perp}$, which is about $1$ for $\omega\sim\omega_{ci}$. This result shows very weak dependance on $\omega_{ce}/\omega_{pe}$ in this frequency range. 
  
\begin{figure}
\begin{center}
\includegraphics[]{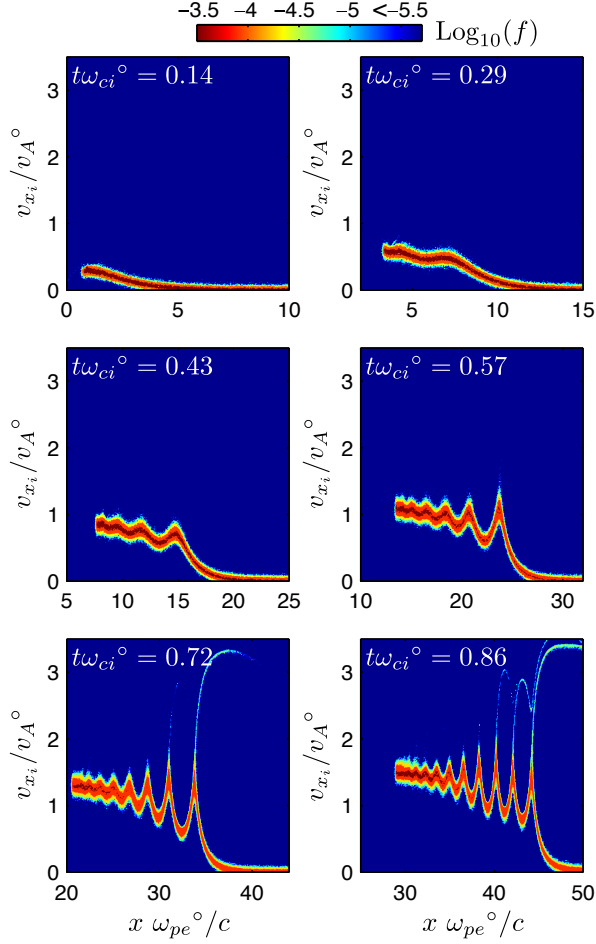}
\caption{Close-up on the ion phase-space ($x,v_x$) density distribution at various times. The ion velocity $v_{x_i}$ and the axial position $x$ are normalized by respectively the initial Alfven velocity ${v_A}^{\circ}$ and the electron skin depth $c/{\omega_{pe}}^{\circ}$. The shock is moving towards the right.  Reflected ions, which are characteristic of a super-critical shock, are clearly seen for $t{\omega_{ci}}^{\circ} = 0.72$ and $0.86$, with $v_{x_i}\sim 3.4~{v_A}^{\circ}$. Initial ion thermal velocity $(k_B {T_i}^{\circ}/m_p)^{1/2}\sim {v_A}^{\circ}/18$.}
\label{Fig:Ion_phase_space}
\end{center}
\end{figure}

For larger driving field amplitudes, $B/{B}^{\circ}-1\sim \mathcal{O}(1)$, features typical of a dispersive shock wave~\cite{Sagdeev1966,Biskamp1973} develop. Similarly to what is observed for finite-amplitude waves propagating perpendicular to a background magnetic field~\cite{Adlam1960,Auer1961,Morton1962,Rossow1965}, a sharp leading edge followed by a trailing wave train downstream of the shock is seen Fig.~\ref{Fig:BFieldContours}. Consistent with these results~\cite{Auer1961}, both the amplitude and wavelength of the pulses forming these compression waves decrease away from the leading edge. Concurrently, space charge separation grows and the longitudinal electric field becomes stronger~\cite{Adlam1958,Ohsawa1985,Ohsawa1985a,Ohsawa1986,Rau1998}. This manifests as an increase of the $x$-component of the ion velocity, as shown in the first four frames in Fig.~\ref{Fig:Ion_phase_space} ($0\leq t{\omega_{ci}}^{\circ}\leq0.6$).

As the amplitude of the driving field grows larger, the shock or discontinuity becomes super-critical. By super-critical we mean that the ions are reflected by the shock, following, for example, the identification of such a transition in Refs ~\cite{Biskamp1973} and \cite{Balogh2013}. Reflection of some of the upstream ions by the shock is then an important mechanism for shock dissipation~\cite{Balogh2013}. Evidence of ion reflection is found in the last two frames in Fig.~\ref{Fig:Ion_phase_space} ($t{\omega_{ci}}^{\circ} = 0.72$ and $0.86$). Upstream ions are reflected by the potential well associated to the shock, and move ahead of the shock with a velocity $v_{x_i}\sim2 v_s$, where $v_s$ is the shock velocity in the laboratory frame. An increase of the ion temperature is observed both upstream and downstream of the shock. Downstream of the shock, in the piston region, ion trapping such as observed in the last frame in Fig.~\ref{Fig:Ion_phase_space}, provides additional dissipation. Although the trailing wave train persists in the super-critical regime, as expected for low $\beta$ plasmas~\cite{Biskamp1973}, it is partly damped as a result of ion trapping.

An estimate of the driving to bias field ratio leading to the formation of a super-critical shock wave can be inferred from this data set. Taking $t{\omega_{ci}}^{\circ} = 0.7$ as the onset, one gets $(B/{B}^{\circ}-1)\sim1.7$. At this instant, the velocity of the accelerating shock front inferred from simulations is $v_s\sim 1.7~{v_A}^{\circ}$, so that the critical Mach number $\mathcal{M}_c \sim1.7$. This value is as expected lower than the $\mathcal{M}_c = 2.76$ typically quoted for a resistive shock propagating perpendicularly into a cold plasma~\cite{Marshall1955}, but close to the well-known limit $\mathcal{M}_c = 2$ for magnetosonic soliton obtained for vanishing resistivity~\cite{Adlam1958,Davis1958,Auer1961,Biskamp1973,Ohsawa2012}. The difference with the $\mathcal{M}_c = 2$ limit might be explained by the finite temperature plasma~\cite{Auer1971} upstream of the shock. 

%As seen in Fig~\ref{Fig:Foot}, $B/n$ remains on average close to ${B}^{\circ}/{n}^{\circ}$, and hence the Alfven velocity in the piston region behind the shock scales roughly as ${v_A}^{\circ}\sqrt{B/B^{\circ}}$. For $(B/{B}^{\circ}-1)\sim1.7$, this gives $v_A\sim1.6~{v_A}^{\circ}$.

\begin{figure}
\begin{center}
\subfigure[)~Magnetic field $B$\label{Fig:BFieldFoot}]{\includegraphics[]{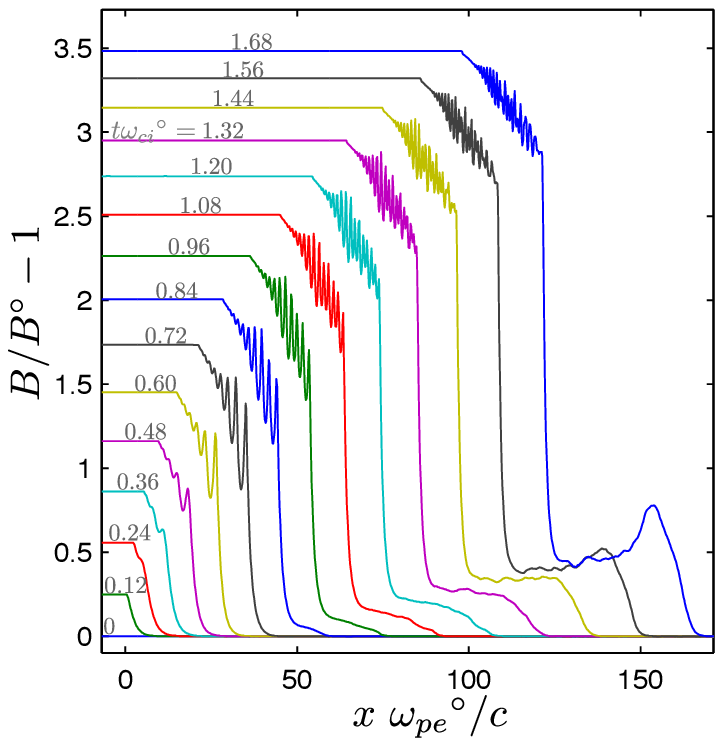}}\hspace{0.5cm}\subfigure[)~Ion density $n_i$\label{Fig:NiFoot}]{\includegraphics[]{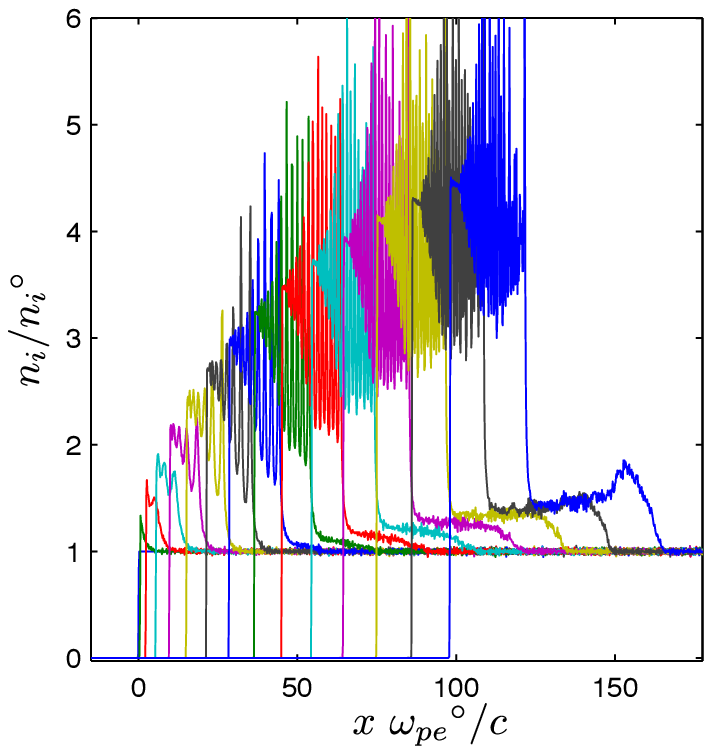}}
\caption{Spatial profile of the magnetic field (\subref{Fig:BFieldFoot}) and ion density (\subref{Fig:NiFoot}) for different values of $t{\omega_{ci}}^{\circ}$. The foot formation upstream of the shock is clearly visible for  $t{\omega_{ci}}^{\circ}\geq0.75$. the density downstream of the shock increases with $B$, creating a hollow profile. For $t{\omega_{ci}}^{\circ}\geq1.4$, the deflection of reflected ions by the magnetic field is responsible for the formation of a local field maximum upstream of the shock.}
\label{Fig:Foot}
\end{center}
\end{figure}

As the shock grows and propagates, the density downstream of the shock increases with $B$. As seen in Fig~\ref{Fig:Foot}, $B/n$ remains on average close to ${B}^{\circ}/{n}^{\circ}$, and a hollow density profile is formed. The density in the piston region displays large oscillations in response to the compression waves observed in this region. The amplitude of these oscillations increases with $B$ for $t{\omega_{ci}}^{\circ}\lesssim 1.2$, and then seems to saturate. This pattern is consistent with the magnetic field profiles in the piston region. In addition, the onset of this saturation phase coincides with the development of turbulent features in the magnetic field profiles downstream of the shock, which are typical of a laminar to turbulent shock transition~\cite{Tidman1971}. The wavelength $\lambda$ of the density oscillations decreases with $B$, with $0.6\lesssim\lambda {\omega_{pe}}^{\circ}/c\lesssim 3$ in Fig.~\ref{Fig:Foot}, which roughly gives $\lambda \propto {B}^{\circ}/B$. Interestingly, since ${\omega_{ci}}^{\circ}{\omega_{ce}}^{\circ}\ll{{\omega_{pi}}^{\circ}}^2$, a $B^{-1}$ scaling corresponds in this case to $1/\omega_{lh}$. This wavelength decrease with a driving field amplitude increase is consistent with two-fluid modeling results~\cite{Morton1962}.

In addition to ion reflection, another typical feature of super-critical shocks is the presence of a foot in the magnetic field profile upstream of the shock~\cite{Woods1971,Biskamp1973}. This foot is due to the reflection of the upstream ions by the potential hill in the shock, and its extension increases as the reflected ions propagate ahead of the shock. The formation of this foot in the magnetic field profile is clearly seen in Fig.~\ref{Fig:BFieldFoot}, and its onset for $t{\omega_{ci}}^{\circ}\sim 0.75$ is quite consistent with the reflected ions identified in Fig.~\ref{Fig:Ion_phase_space}. However, although the extension of the foot does initially increase as the reflected ions propagate ahead of the shock, this mechanism diminishes progressively due to the deflection of these ions by the magnetic field. For $t{\omega_{ci}}^{\circ}\geq1.4$, the foot extension does not vary significantly, and the magnetic field increases ($\partial B/\partial x>0$) due to the negative transverse ion current. This pattern develops until the two counter-propagating beams ahead of their respective shock collide. 

\section{Peak-compression and expansion}

\label{Sec:V}

\begin{figure}
\begin{center}
\subfigure[)~$t{\omega_{ci}}^{\circ}\sim1.87$ ]{\includegraphics[width=8cm]{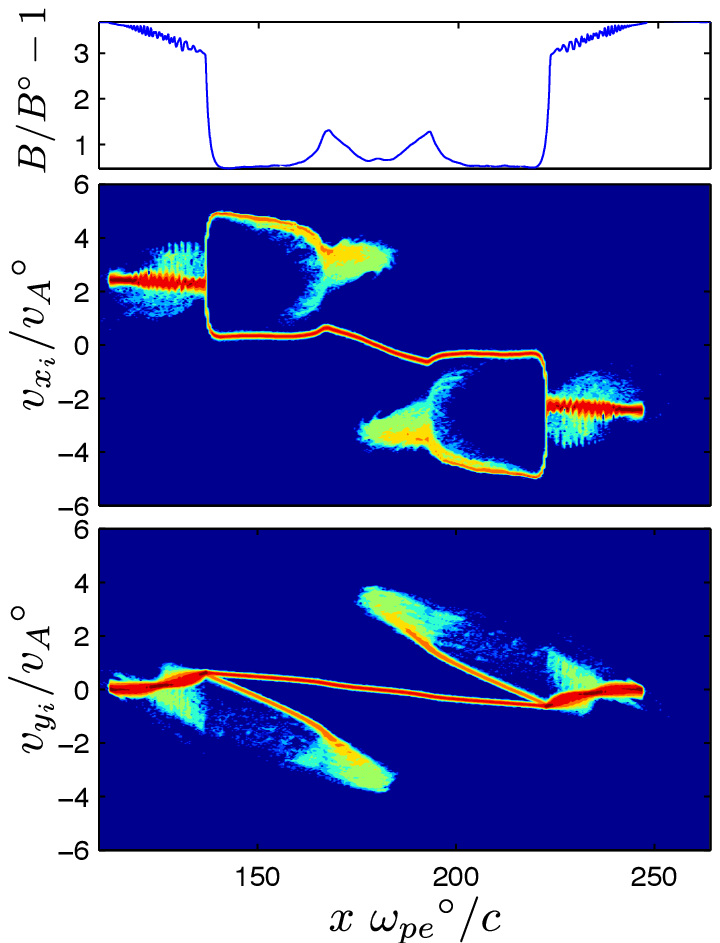}\label{Fig:Phase_space_14}}\subfigure[)~$t{\omega_{ci}}^{\circ}\sim2.01$ ]{\includegraphics[width=8cm]{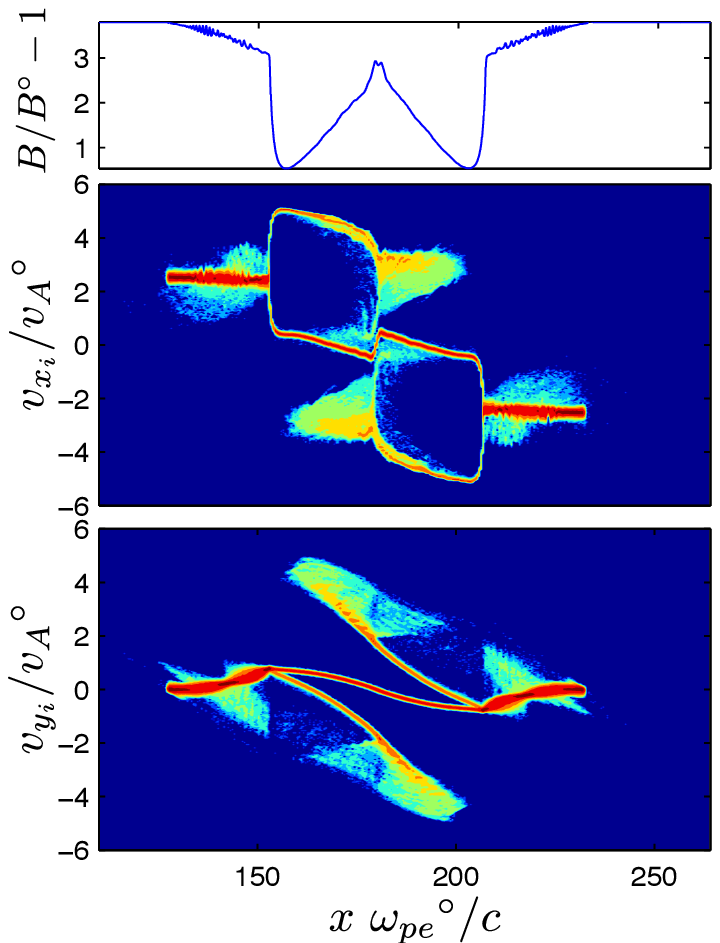}\label{Fig:Phase_space_15}}\\\subfigure[)~$t{\omega_{ci}}^{\circ}\sim2.16$]{\includegraphics[width=8cm]{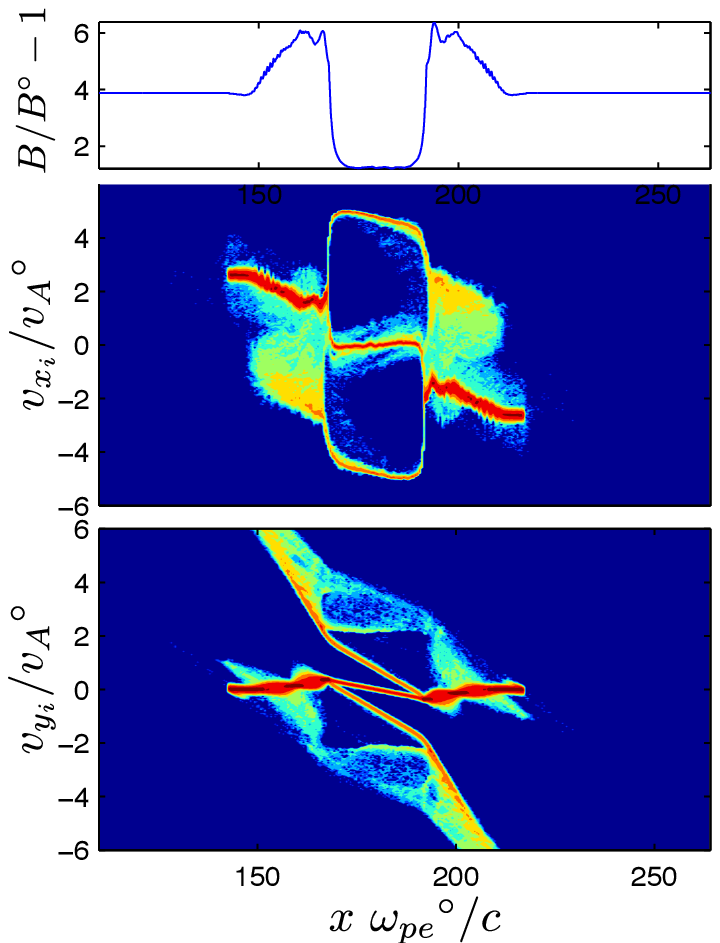}\label{Fig:Phase_space_16}}\subfigure[)~$t{\omega_{ci}}^{\circ}\sim2.30$]{\includegraphics[width=8cm]{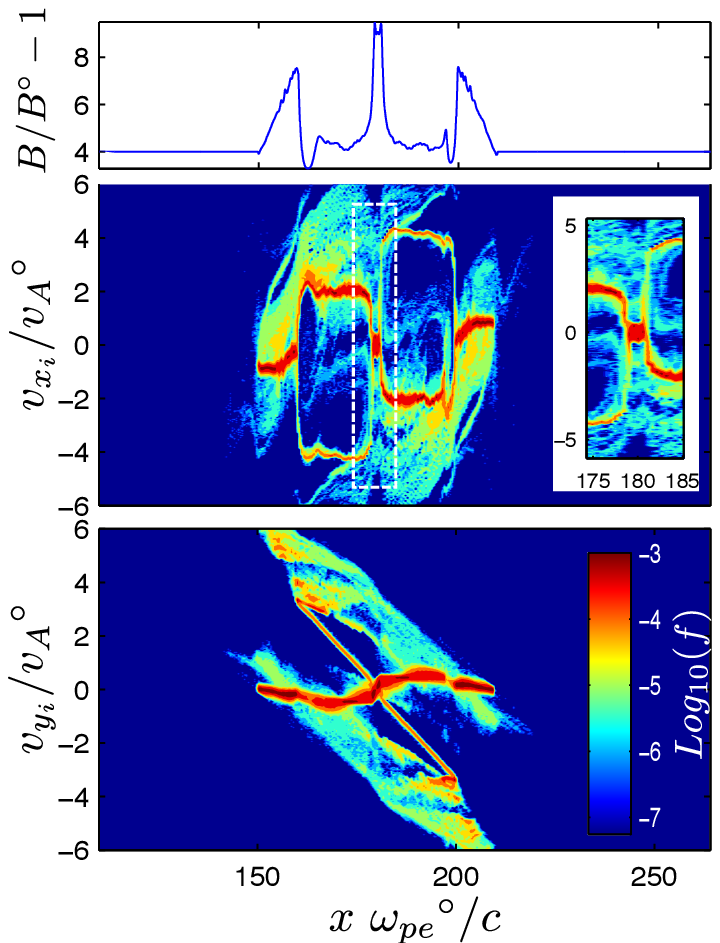}\label{Fig:Phase_space_17}}
\label{Fig:Phase_space}
\caption{Ion phase space and magnetic field profile evolution around peak compression: prior to~(\subref{Fig:Phase_space_14}) and immediately after~(\subref{Fig:Phase_space_15}) the interaction of the counter-propagating ion beams, during the interaction of the ions beam and the counter-propagating shock~(\subref{Fig:Phase_space_16}) and after the interaction of the two counter-propagating shocks~(\subref{Fig:Phase_space_17}). The close up in the second panel in Fig.~(\subref{Fig:Phase_space_17}) highlights the slowing down of piston ions by the counter-propagating shock.}
\end{center}
\end{figure}

When the two counter-propagating ion beams reach the mid-plane of the simulation domain, which happens for $t{\omega_{ci}}^{\circ}\sim1.85$, the magnetic field carried by each of the feet add and the magnetic field in the mid-plane grows momentarily, with $B\sim5{B}^{\circ}$. This is seen is Fig.~\ref{Fig:Phase_space_14} and \ref{Fig:Phase_space_15}. The magnetic field in the mid-plane then decreases to $B\sim2{B}^{\circ}$ as the head of the beams move further away from the mid-plane. 

Next, when the right (\emph{resp.} left) propagating beam encounters the left (\emph{resp.} right) shock, the beam ions loose momentum in the $x$ direction (Fig.~\ref{Fig:Phase_space_16}). Simultaneously, magnetic deflection of the beam ion is amplified due to the increase of $B$ across the shock. As a result, the amplitude of the negative (\emph{resp.} positive) ion transverse current $j_{y_i}$ increases, and the magnetic field downstream of each shock increases beyond the driving field intensity. Progressive magnetic deflection of the ions in the piston region amplifies this response. 

Later on, when the two counter-propagating shocks collide, part of the ions located in the piston (downstream) region of one shock go through the other shock. In the process, these ions loose most of their momentum along the $x$ direction. This is illustrated by the central region highlighted in the close-up in the second panel in Fig.~\ref{Fig:Phase_space_17}. The rest of the ions from the piston region are reflected by the incoming shock. As a result, the velocity of these ions in the laboratory frame is $v_x\sim-v_p-v_s$, where $v_p$ is the ion velocity in the piston region and $v_s\sim $ is the velocity of the incoming shock. In the laboratory frame, the left moving shock velocity is $v_s \lesssim -2{v_A}^{\circ}$ and the right moving piston ions velocity is $v_p \lesssim 2{v_A}^{\circ}$, so that $v_x\sim -4{v_A}^{\circ}$. This left propagating ion beam and the corresponding right propagating ion beam produced by the right moving shock  are seen respectively in the left ($v_x\sim-4{v_A}^{\circ}$) and right ($v_x\sim4{v_A}^{\circ}$) half of the domain in the second panel in Fig.~\ref{Fig:Phase_space_17}. 

Consider now the magnetic field profile in the first panel in Fig.~\ref{Fig:Phase_space_17}. The magnetic field peaks in the mid-plane immediately after the two shocks collide. This strong and localized increase is due to the positive (\emph{resp.} negative) transverse electron current $j_{y_e}$ associated with the positive (\emph{resp.} negative) longitudinal electric field $E_x$ of the right (\emph{resp.} left) propagating shock. Prior to the shocks colliding, these electron currents are responsible for the magnetic field drop in the central region. Immediately after the shocks propagate past the mid-plane, these currents lead to a strong field amplification in the region in between the shocks. 

The instant when shocks reach the mid-plane corresponds as well to peak compression. For $t {\omega_{ci}}^{\circ} \sim 2.26$, the plasma region width is $\Delta x\sim 62 c/{\omega_{pe}}^{\circ}$, which is about ${\Delta x}^{\circ}/6$, with ${\Delta x}^{\circ}$ the slab width at $t=0$. This compression value first seems surprising since one would expect ${\Delta x}^{\circ}/\Delta x \sim B/{B}^{\circ} \sim 2\delta_B+1$, which is equal to $5$ for the conditions studied here (see Table~\ref{Table:compression_parameters}). However, a more careful look at the magnetic field profile provides an explanation for this result. Prior to the shocks colliding, the magnetic field in the outer plasma region is larger than the driving field due to the ion beam-shock interaction described above. This is illustrated in the first panel in Fig.~\ref{Fig:Phase_space_16}. Since $B/n$ remains nearly constant, the compression $n/{n}^{\circ}$ in this region is greater than $2\delta_B+1$, while $n/{n}^{\circ}\sim 2\delta_B+1$ in the central region as the shocks collide. Consequently, the peak compression is greater than what one would expect considering only the driving to bias field ratio. 

Past peak compression, the plasma expands again. Concurrently, the magnetic field peak in the mid-plane decreases and broadens. Eventually, the magnetic field profile returns to a hollow profile, with $B\sim(2\delta_B+1){B}^{\circ}$ outside of the plasma region, and $B\leq(2\delta_B+1){B}^{\circ}$ in the plasma region.

\section{Summary}
\label{Sec:Summary}

The practicality of plasma density control through fast magnetic compression was studied using fully electromagnetic particle in cell simulations. In particular, the feasibility of plasma densification for wave properties control, such as envisioned for electron dephasing suppression in plasma-based particle accelerators, was investigated.  

In the plasma regime relevant to this application, namely compression on timescales shorter than the ion gyro-period, simulation results obtained in a slab geometry indicate a plasma dynamic significantly different from the homogeneous compression initially proposed. During the compression phase, the plasma density is seen to be strongly non-uniform, and exhibits a hollow profile. Density variations as large as the driving to bias field ratio times the initial density are observed over a distance on the order of an electron skin depth. Furthermore, collisionless oscillatory behavior, with multiple successive compression and expansion phases, similarly to hydromagnetic oscillations, is observed. As the plasma is compressed, strong plasma heating is also observed, with the largest part of the kinetic energy in the ion motion. Simple estimates suggest $\beta\sim\mathcal{O}(1)$ near peak compression.

Instead of the uniform plasma densification initially suggested, fast magnetic compression is shown to take the form of two counter-propagating and accelerating shock waves. Downstream of each shock, in the piston region, the density increases with the driving field, while the plasma upstream of the shock remains initially unperturbed. As the driving magnetic field grows, the shock becomes supercritical, and an ion beam forms ahead of the shock, as some of the incident upstream ions are reflected. The deflection of this ion beam by the magnetic field is suddenly amplified as the head of the beam interacts with the counter-propagating shock, leading temporarily to a magnetic field greater than the driving field in the plasma. Peak compression coincides with the collision of the two counter-propagating shocks. In the process, a fraction of ions downstream of a given shock is reflected by the counter-propagating shock.

The phenomena uncovered in this paper are expected to take place for fast compression experiments in which: (\emph{i}) the driving to bias field ratio $\delta_B\gtrsim2$; (\emph{ii}) the compression time $\tau_r$ is shorter than the ion gyro-period; (\emph{iii}) the compression time is shorter than the system size $\Delta x$ divided by the Alfven speed $V_a$; and (\emph{iv}) the compression time is shorter than the ion and electron collision times. For very large systems, defined as $\tau_r V_A \ll \Delta x$, the dynamics of the shock foot is expected to differ since in this limit beam ions will be turned around by the magnetic field, and possibly re-interact with the shock, before interacting with the counter-propagating shock. Furthermore, space charge effects are expected to become stronger for regimes where the plasma frequency becomes comparable, or even smaller, than the electron gyro-frequency.

Although the results presented here indicate a densification mechanism much more complex than what was initially envisioned for wave properties control in plasma-based particle accelerators, these results are not necessarily incompatible with this application.In fact, with the transit time of a light wave being much shorter than the compression time, it might be possible to take advantage of a specific plasma density profile resulting from shock compression at a given time. It remains, however, to evaluate whether these benefits balance the added complexity associated with time-resolved shock compression.

\section*{Acknowledgments}
The authors would like to thank Amnon Fruchtman for constructive discussions.

This work was supported by DTRA Grant No. HDTRA1-11-1-0037 and by DOE Grant No. 67350-9960 (Prime No. DOE DE-NA0001836).

%\bibliography{aipsamp}
%\bibliography{References}

\begin{thebibliography}{28}%
\makeatletter
\providecommand \@ifxundefined [1]{%
 \@ifx{#1\undefined}
}%
\providecommand \@ifnum [1]{%
 \ifnum #1\expandafter \@firstoftwo
 \else \expandafter \@secondoftwo
 \fi
}%
\providecommand \@ifx [1]{%
 \ifx #1\expandafter \@firstoftwo
 \else \expandafter \@secondoftwo
 \fi
}%
\providecommand \natexlab [1]{#1}%
\providecommand \enquote  [1]{``#1''}%
\providecommand \bibnamefont  [1]{#1}%
\providecommand \bibfnamefont [1]{#1}%
\providecommand \citenamefont [1]{#1}%
\providecommand \href@noop [0]{\@secondoftwo}%
\providecommand \href [0]{\begingroup \@sanitize@url \@href}%
\providecommand \@href[1]{\@@startlink{#1}\@@href}%
\providecommand \@@href[1]{\endgroup#1\@@endlink}%
\providecommand \@sanitize@url [0]{\catcode `\\12\catcode `\$12\catcode
  `\&12\catcode `\#12\catcode `\^12\catcode `\_12\catcode `\%12\relax}%
\providecommand \@@startlink[1]{}%
\providecommand \@@endlink[0]{}%
\providecommand \url  [0]{\begingroup\@sanitize@url \@url }%
\providecommand \@url [1]{\endgroup\@href {#1}{\urlprefix }}%
\providecommand \urlprefix  [0]{URL }%
\providecommand \Eprint [0]{\href }%
\providecommand \doibase [0]{http://dx.doi.org/}%
\providecommand \selectlanguage [0]{\@gobble}%
\providecommand \bibinfo  [0]{\@secondoftwo}%
\providecommand \bibfield  [0]{\@secondoftwo}%
\providecommand \translation [1]{[#1]}%
\providecommand \BibitemOpen [0]{}%
\providecommand \bibitemStop [0]{}%
\providecommand \bibitemNoStop [0]{.\EOS\space}%
\providecommand \EOS [0]{\spacefactor3000\relax}%
\providecommand \BibitemShut  [1]{\csname bibitem#1\endcsname}%
\let\auto@bib@innerbib\@empty
%</preamble>
\bibitem [{\citenamefont {Schmit}\ and\ \citenamefont
  {Fisch}(2012)}]{Schmit2012}%
  \BibitemOpen
  \bibfield  {author} {\bibinfo {author} {\bibfnamefont {P.~F.}\ \bibnamefont
  {Schmit}}\ and\ \bibinfo {author} {\bibfnamefont {N.~J.}\ \bibnamefont
  {Fisch}},\ }\href {\doibase 10.1103/PhysRevLett.109.255003} {\bibfield
  {journal} {\bibinfo  {journal} {Phys. Rev. Lett.}\ }\textbf {\bibinfo
  {volume} {109}},\ \bibinfo {pages} {255003} (\bibinfo {year}
  {2012})}\BibitemShut {NoStop}%
\bibitem [{\citenamefont {Esarey}, \citenamefont {Schroeder},\ and\
  \citenamefont {Leemans}(2009)}]{Esarey2009}%
  \BibitemOpen
  \bibfield  {author} {\bibinfo {author} {\bibfnamefont {E.}~\bibnamefont
  {Esarey}}, \bibinfo {author} {\bibfnamefont {C.~B.}\ \bibnamefont
  {Schroeder}}, \ and\ \bibinfo {author} {\bibfnamefont {W.~P.}\ \bibnamefont
  {Leemans}},\ }\href {\doibase 10.1103/RevModPhys.81.1229} {\bibfield
  {journal} {\bibinfo  {journal} {Rev. Mod. Phys.}\ }\textbf {\bibinfo {volume}
  {81}},\ \bibinfo {pages} {1229} (\bibinfo {year} {2009})}\BibitemShut
  {NoStop}%
\bibitem [{\citenamefont {Morse}(1967)}]{Morse1967}%
  \BibitemOpen
  \bibfield  {author} {\bibinfo {author} {\bibfnamefont {R.~L.}\ \bibnamefont
  {Morse}},\ }\href {\doibase 10.1063/1.1762215} {\bibfield  {journal}
  {\bibinfo  {journal} {Physics of Fluids}\ }\textbf {\bibinfo {volume} {10}},\
  \bibinfo {pages} {1017} (\bibinfo {year} {1967})}\BibitemShut {NoStop}%
\bibitem [{\citenamefont {Dove}(1971)}]{Dove1971}%
  \BibitemOpen
  \bibfield  {author} {\bibinfo {author} {\bibfnamefont {W.~F.}\ \bibnamefont
  {Dove}},\ }\href {\doibase 10.1063/1.1693341} {\bibfield  {journal} {\bibinfo
   {journal} {Physics of Fluids}\ }\textbf {\bibinfo {volume} {14}},\ \bibinfo
  {pages} {2359} (\bibinfo {year} {1971})}\BibitemShut {NoStop}%
\bibitem [{\citenamefont {Rosenbluth}(1954)}]{Rosenbluth1954}%
  \BibitemOpen
  \bibfield  {author} {\bibinfo {author} {\bibfnamefont {M.}~\bibnamefont
  {Rosenbluth}},\ }\href@noop {} {\enquote {\bibinfo {title} {Infinite
  conductivity theory of the pinch},}\ }\bibinfo {type} {Tech. Rep.}\ \bibinfo
  {number} {LA 1850}\ (\bibinfo  {institution} {Los Alamos Laboratory},\
  \bibinfo {year} {1954})\BibitemShut {NoStop}%
\bibitem [{\citenamefont {Uchida}, \citenamefont {Sato},\ and\ \citenamefont
  {Hamada}(1962)}]{Uchida1962}%
  \BibitemOpen
  \bibfield  {author} {\bibinfo {author} {\bibfnamefont {T.}~\bibnamefont
  {Uchida}}, \bibinfo {author} {\bibfnamefont {M.}~\bibnamefont {Sato}}, \ and\
  \bibinfo {author} {\bibfnamefont {S.}~\bibnamefont {Hamada}},\ }\href
  {\doibase 10.1088/0029-5515/2/1-2/010} {\bibfield  {journal} {\bibinfo
  {journal} {Nuclear Fusion}\ }\textbf {\bibinfo {volume} {2}},\ \bibinfo
  {pages} {70} (\bibinfo {year} {1962})}\BibitemShut {NoStop}%
\bibitem [{\citenamefont {Arber}\ \emph {et~al.}(2015)\citenamefont {Arber},
  \citenamefont {Bennett}, \citenamefont {Brady}, \citenamefont
  {Lawrence-Douglas}, \citenamefont {Ramsay}, \citenamefont {Sircombe},
  \citenamefont {Gillies}, \citenamefont {Evans}, \citenamefont {Schmitz},
  \citenamefont {Bell},\ and\ \citenamefont {Ridgers}}]{Arber2015}%
  \BibitemOpen
  \bibfield  {author} {\bibinfo {author} {\bibfnamefont {T.~D.}\ \bibnamefont
  {Arber}}, \bibinfo {author} {\bibfnamefont {K.}~\bibnamefont {Bennett}},
  \bibinfo {author} {\bibfnamefont {C.~S.}\ \bibnamefont {Brady}}, \bibinfo
  {author} {\bibfnamefont {A.}~\bibnamefont {Lawrence-Douglas}}, \bibinfo
  {author} {\bibfnamefont {M.~G.}\ \bibnamefont {Ramsay}}, \bibinfo {author}
  {\bibfnamefont {N.~J.}\ \bibnamefont {Sircombe}}, \bibinfo {author}
  {\bibfnamefont {P.}~\bibnamefont {Gillies}}, \bibinfo {author} {\bibfnamefont
  {R.~G.}\ \bibnamefont {Evans}}, \bibinfo {author} {\bibfnamefont
  {H.}~\bibnamefont {Schmitz}}, \bibinfo {author} {\bibfnamefont {A.~R.}\
  \bibnamefont {Bell}}, \ and\ \bibinfo {author} {\bibfnamefont {C.~P.}\
  \bibnamefont {Ridgers}},\ }\href {\doibase 10.1088/0741-3335/57/11/113001}
  {\bibfield  {journal} {\bibinfo  {journal} {Plasma Physics and Controlled
  Fusion}\ }\textbf {\bibinfo {volume} {57}},\ \bibinfo {pages} {113001}
  (\bibinfo {year} {2015})}\BibitemShut {NoStop}%
\bibitem [{\citenamefont {Niblett}\ and\ \citenamefont
  {Green}(1959)}]{Niblett1959}%
  \BibitemOpen
  \bibfield  {author} {\bibinfo {author} {\bibfnamefont {G.~B.~F.}\
  \bibnamefont {Niblett}}\ and\ \bibinfo {author} {\bibfnamefont {T.~S.}\
  \bibnamefont {Green}},\ }\href {\doibase 10.1088/0370-1328/74/6/311}
  {\bibfield  {journal} {\bibinfo  {journal} {Proceedings of the Physical
  Society}\ }\textbf {\bibinfo {volume} {74}},\ \bibinfo {pages} {737}
  (\bibinfo {year} {1959})}\BibitemShut {NoStop}%
\bibitem [{\citenamefont {Fisher}, \citenamefont {Green},\ and\ \citenamefont
  {Niblett}(1962)}]{Fisher1962}%
  \BibitemOpen
  \bibfield  {author} {\bibinfo {author} {\bibfnamefont {D.~L.}\ \bibnamefont
  {Fisher}}, \bibinfo {author} {\bibfnamefont {T.~S.}\ \bibnamefont {Green}}, \
  and\ \bibinfo {author} {\bibfnamefont {G.~B.~F.}\ \bibnamefont {Niblett}},\
  }\href {\doibase 10.1088/0368-3281/4/3/405} {\bibfield  {journal} {\bibinfo
  {journal} {Journal of Nuclear Energy. Part C, Plasma Physics, Accelerators,
  Thermonuclear Research}\ }\textbf {\bibinfo {volume} {4}},\ \bibinfo {pages}
  {181} (\bibinfo {year} {1962})}\BibitemShut {NoStop}%
\bibitem [{\citenamefont {Martone}\ and\ \citenamefont
  {Segre}(1971)}]{Martone1971}%
  \BibitemOpen
  \bibfield  {author} {\bibinfo {author} {\bibfnamefont {M.}~\bibnamefont
  {Martone}}\ and\ \bibinfo {author} {\bibfnamefont {S.~E.}\ \bibnamefont
  {Segre}},\ }\href {\doibase 10.1088/0032-1028/13/2/007} {\bibfield  {journal}
  {\bibinfo  {journal} {Plasma Physics}\ }\textbf {\bibinfo {volume} {13}},\
  \bibinfo {pages} {173} (\bibinfo {year} {1971})}\BibitemShut {NoStop}%
\bibitem [{\citenamefont {Sagdeev}(1966)}]{Sagdeev1966}%
  \BibitemOpen
  \bibfield  {author} {\bibinfo {author} {\bibfnamefont {R.~Z.}\ \bibnamefont
  {Sagdeev}},\ }\href@noop {} {\bibfield  {journal} {\bibinfo  {journal} {Rev.
  Plasma Physics}\ }\textbf {\bibinfo {volume} {4}},\ \bibinfo {pages} {23}
  (\bibinfo {year} {1966})}\BibitemShut {NoStop}%
\bibitem [{\citenamefont {Biskamp}(1973)}]{Biskamp1973}%
  \BibitemOpen
  \bibfield  {author} {\bibinfo {author} {\bibfnamefont {D.}~\bibnamefont
  {Biskamp}},\ }\href {\doibase 10.1088/0029-5515/13/5/010} {\bibfield
  {journal} {\bibinfo  {journal} {Nuclear Fusion}\ }\textbf {\bibinfo {volume}
  {13}},\ \bibinfo {pages} {719} (\bibinfo {year} {1973})}\BibitemShut
  {NoStop}%
\bibitem [{\citenamefont {Adlam}\ and\ \citenamefont
  {Allen}(1960)}]{Adlam1960}%
  \BibitemOpen
  \bibfield  {author} {\bibinfo {author} {\bibfnamefont {J.~H.}\ \bibnamefont
  {Adlam}}\ and\ \bibinfo {author} {\bibfnamefont {J.~E.}\ \bibnamefont
  {Allen}},\ }\href {\doibase 10.1088/0370-1328/75/5/302} {\bibfield  {journal}
  {\bibinfo  {journal} {Proceedings of the Physical Society}\ }\textbf
  {\bibinfo {volume} {75}},\ \bibinfo {pages} {640} (\bibinfo {year}
  {1960})}\BibitemShut {NoStop}%
\bibitem [{\citenamefont {Auer}, \citenamefont {Hurwitz},\ and\ \citenamefont
  {Kilb}(1961)}]{Auer1961}%
  \BibitemOpen
  \bibfield  {author} {\bibinfo {author} {\bibfnamefont {P.~L.}\ \bibnamefont
  {Auer}}, \bibinfo {author} {\bibfnamefont {H.}~\bibnamefont {Hurwitz}}, \
  and\ \bibinfo {author} {\bibfnamefont {R.~W.}\ \bibnamefont {Kilb}},\ }\href
  {\doibase 10.1063/1.1706455} {\bibfield  {journal} {\bibinfo  {journal}
  {Physics of Fluids}\ }\textbf {\bibinfo {volume} {4}},\ \bibinfo {pages}
  {1105} (\bibinfo {year} {1961})}\BibitemShut {NoStop}%
\bibitem [{\citenamefont {Morton}(1962)}]{Morton1962}%
  \BibitemOpen
  \bibfield  {author} {\bibinfo {author} {\bibfnamefont {K.~W.}\ \bibnamefont
  {Morton}},\ }\href {\doibase 10.1017/S0022112062001299} {\bibfield  {journal}
  {\bibinfo  {journal} {Journal of Fluid Mechanics}\ }\textbf {\bibinfo
  {volume} {14}},\ \bibinfo {pages} {369} (\bibinfo {year} {1962})}\BibitemShut
  {NoStop}%
\bibitem [{\citenamefont {Rossow}(1965)}]{Rossow1965}%
  \BibitemOpen
  \bibfield  {author} {\bibinfo {author} {\bibfnamefont {V.~J.}\ \bibnamefont
  {Rossow}},\ }\href {\doibase 10.1063/1.1761230} {\bibfield  {journal}
  {\bibinfo  {journal} {Physics of Fluids}\ }\textbf {\bibinfo {volume} {8}},\
  \bibinfo {pages} {358} (\bibinfo {year} {1965})}\BibitemShut {NoStop}%
\bibitem [{\citenamefont {Adlam}\ and\ \citenamefont
  {Allen}(1958)}]{Adlam1958}%
  \BibitemOpen
  \bibfield  {author} {\bibinfo {author} {\bibfnamefont {J.~H.}\ \bibnamefont
  {Adlam}}\ and\ \bibinfo {author} {\bibfnamefont {J.~E.}\ \bibnamefont
  {Allen}},\ }\href {\doibase 10.1080/14786435808244566} {\bibfield  {journal}
  {\bibinfo  {journal} {Philosophical Magazine}\ }\textbf {\bibinfo {volume}
  {3}},\ \bibinfo {pages} {448} (\bibinfo {year} {1958})}\BibitemShut {NoStop}%
\bibitem [{\citenamefont {Ohsawa}(1985{\natexlab{a}})}]{Ohsawa1985}%
  \BibitemOpen
  \bibfield  {author} {\bibinfo {author} {\bibfnamefont {Y.}~\bibnamefont
  {Ohsawa}},\ }\href {\doibase 10.1063/1.865394} {\bibfield  {journal}
  {\bibinfo  {journal} {Physics of Fluids}\ }\textbf {\bibinfo {volume} {28}},\
  \bibinfo {pages} {2130} (\bibinfo {year} {1985}{\natexlab{a}})}\BibitemShut
  {NoStop}%
\bibitem [{\citenamefont {Ohsawa}(1985{\natexlab{b}})}]{Ohsawa1985a}%
  \BibitemOpen
  \bibfield  {author} {\bibinfo {author} {\bibfnamefont {Y.}~\bibnamefont
  {Ohsawa}},\ }\href {\doibase 10.1143/JPSJ.54.1657} {\bibfield  {journal}
  {\bibinfo  {journal} {J. Phys. Soc. Jpn.}\ }\textbf {\bibinfo {volume}
  {54}},\ \bibinfo {pages} {1657} (\bibinfo {year}
  {1985}{\natexlab{b}})}\BibitemShut {NoStop}%
\bibitem [{\citenamefont {Ohsawa}(1986)}]{Ohsawa1986}%
  \BibitemOpen
  \bibfield  {author} {\bibinfo {author} {\bibfnamefont {Y.}~\bibnamefont
  {Ohsawa}},\ }\href {\doibase 10.1063/1.865932} {\bibfield  {journal}
  {\bibinfo  {journal} {Physics of Fluids}\ }\textbf {\bibinfo {volume} {29}},\
  \bibinfo {pages} {773} (\bibinfo {year} {1986})}\BibitemShut {NoStop}%
\bibitem [{\citenamefont {Rau}\ and\ \citenamefont {Tajima}(1998)}]{Rau1998}%
  \BibitemOpen
  \bibfield  {author} {\bibinfo {author} {\bibfnamefont {B.}~\bibnamefont
  {Rau}}\ and\ \bibinfo {author} {\bibfnamefont {T.}~\bibnamefont {Tajima}},\
  }\href {\doibase 10.1063/1.873076} {\bibfield  {journal} {\bibinfo  {journal}
  {Physics of Plasmas}\ }\textbf {\bibinfo {volume} {5}},\ \bibinfo {pages}
  {3575} (\bibinfo {year} {1998})}\BibitemShut {NoStop}%
\bibitem [{\citenamefont {Balogh}\ and\ \citenamefont
  {Treumann}(2013)}]{Balogh2013}%
  \BibitemOpen
  \bibfield  {author} {\bibinfo {author} {\bibfnamefont {A.}~\bibnamefont
  {Balogh}}\ and\ \bibinfo {author} {\bibfnamefont {R.~A.}\ \bibnamefont
  {Treumann}},\ }\href {\doibase 10.1007/978-1-4614-6099-2} {\emph {\bibinfo
  {title} {Physics of Collisionless Shocks}}},\ \bibinfo {series} {ISSI
  Scientific Report Series}, Vol.~\bibinfo {volume} {12}\ (\bibinfo
  {publisher} {Springer-Verlag New York},\ \bibinfo {year} {2013})\BibitemShut
  {NoStop}%
\bibitem [{\citenamefont {Marshall}(1955)}]{Marshall1955}%
  \BibitemOpen
  \bibfield  {author} {\bibinfo {author} {\bibfnamefont {W.}~\bibnamefont
  {Marshall}},\ }\href {\doibase 10.1098/rspa.1955.0272} {\bibfield  {journal}
  {\bibinfo  {journal} {Proceedings of the Royal Society of London A:
  Mathematical, Physical and Engineering Sciences}\ }\textbf {\bibinfo {volume}
  {233}},\ \bibinfo {pages} {367} (\bibinfo {year} {1955})}\BibitemShut
  {NoStop}%
\bibitem [{\citenamefont {Davis}, \citenamefont {Lust},\ and\ \citenamefont
  {Schluter}(1958)}]{Davis1958}%
  \BibitemOpen
  \bibfield  {author} {\bibinfo {author} {\bibfnamefont {L.}~\bibnamefont
  {Davis}}, \bibinfo {author} {\bibfnamefont {R.}~\bibnamefont {Lust}}, \ and\
  \bibinfo {author} {\bibfnamefont {A.}~\bibnamefont {Schluter}},\ }\href@noop
  {} {\bibfield  {journal} {\bibinfo  {journal} {Z. Naturf. A}\ }\textbf
  {\bibinfo {volume} {13}},\ \bibinfo {pages} {916} (\bibinfo {year}
  {1958})}\BibitemShut {NoStop}%
\bibitem [{\citenamefont {Ohsawa}(2012)}]{Ohsawa2012}%
  \BibitemOpen
  \bibfield  {author} {\bibinfo {author} {\bibfnamefont {Y.}~\bibnamefont
  {Ohsawa}},\ }\href@noop {} {\enquote {\bibinfo {title} {Ultrarelativistic
  particle acceleration in collisionless shock waves},}\ }\bibinfo {type}
  {Tech. Rep.}\ \bibinfo {number} {IFSR1433}\ (\bibinfo  {institution}
  {Institute for Fusion Studies},\ \bibinfo {year} {2012})\BibitemShut
  {NoStop}%
\bibitem [{\citenamefont {Auer}\ and\ \citenamefont {Evers}(1971)}]{Auer1971}%
  \BibitemOpen
  \bibfield  {author} {\bibinfo {author} {\bibfnamefont {P.~L.}\ \bibnamefont
  {Auer}}\ and\ \bibinfo {author} {\bibfnamefont {W.~H.}\ \bibnamefont
  {Evers}},\ }\href {\doibase 10.1063/1.1693583} {\bibfield  {journal}
  {\bibinfo  {journal} {Physics of Fluids}\ }\textbf {\bibinfo {volume} {14}},\
  \bibinfo {pages} {1177} (\bibinfo {year} {1971})}\BibitemShut {NoStop}%
\bibitem [{\citenamefont {Tidman}\ and\ \citenamefont
  {Krall}(1971)}]{Tidman1971}%
  \BibitemOpen
  \bibfield  {author} {\bibinfo {author} {\bibfnamefont {D.~A.}\ \bibnamefont
  {Tidman}}\ and\ \bibinfo {author} {\bibfnamefont {N.~A.}\ \bibnamefont
  {Krall}},\ }\href@noop {} {\emph {\bibinfo {title} {Shock waves in
  collisionless plasmas}}}\ (\bibinfo  {publisher} {Wiley-Interscience},\
  \bibinfo {year} {1971})\BibitemShut {NoStop}%
\bibitem [{\citenamefont {Woods}(1971)}]{Woods1971}%
  \BibitemOpen
  \bibfield  {author} {\bibinfo {author} {\bibfnamefont {L.~C.}\ \bibnamefont
  {Woods}},\ }\href {\doibase 10.1088/0032-1028/13/4/302} {\bibfield  {journal}
  {\bibinfo  {journal} {Plasma Physics}\ }\textbf {\bibinfo {volume} {13}},\
  \bibinfo {pages} {289} (\bibinfo {year} {1971})}\BibitemShut {NoStop}%
\end{thebibliography}


\end{document}